\documentstyle[aps,epsfig]{revtex}
\def\d{\partial}

\def\m{\mu}
\def\n{\nu}

\def\t{\tau}

\def\~{\tilde}
\def\h{\eta}

\def\bY3{\bar Y_{,3}}
\def\Y3{Y_{,3}}

\def\Y{{\bar Y}}

\def\`{\dot}
\def\be{\begin{equation}}
\def\ee{\end{equation}}
\def\bea{\begin{eqnarray}}
\def\eea{\end{eqnarray}}

\def\fn{\footnote}

\def\mn{{\mu\nu}}

\begin{document}

\title{Orientifold D-String in the Source of the Kerr Spinning Particle}

\author{Alexander Burinskii\\
Gravity Research Group, NSI Russian
Academy of Sciences\\
B. Tulskaya 52, 115191 Moscow, Russia}
\maketitle

\begin{abstract}
\noindent
The model of spinning particle, based on the Kerr-Newman
solution with $|a|>>m$,  is discussed. It is shown that the Kerr singular
ring can be considered as a string with an orientifold world-sheet.
Orientifold adds to the Kerr ring an extra peculiar point, the
fixed point of the world-sheet parity operator $\Omega$.  It is shown that
the Kerr string represents a new type of the folded string solutions
taking the form of the open D-string with joined ends which are in the
circular light-like motion along the Kerr ring.  \end{abstract}

\section{Introduction}

The Kerr rotating black hole solution displays some remarkable relations to
the spinning particles
\cite{Car,Isr,Bur0,IvBur,Lop,BurSen,BurStr,BurSup,Bag}. For parameters of
elementary particles $|a|>> m$, and the black-hole horizons disappear,
obtaining the source in the form of a closed singular ring of the Compton
radius\fn{ $a=J/m$ is the density of angular momentum $J$ per mass $m$. We use
the units $c=\hbar =G=1$, and signature $(-+++)$.}.  In the model of the Kerr
spinning particle \cite{Bur0} this ring was considered as a gravitational
waveguide  containing traveling electromagnetic (and
fermionic) wave excitations.
The assumption that the Kerr singular ring represents a closed
relativistic string was advanced about thirty years ago, \cite{IvBur},
which had confirmations on the level of evidences \cite{BurSen,IvBur1,Bur1}.
However, the
attempts to show it explicitly ran into obstacle which was related with a
very specific motion of the Kerr ring - the light-like sliding along
itself.  It could be described as a string containing the light-like modes of
only one direction.  However, the relativistic string equations do not admit
such solutions.

In this note we resolve this problem showing that the Kerr ring is a
string with orientifold structure, which represents a new type of the
folded string solutions. \fn{Note, that interest to the classical string
solutions,
and especially to the folded ones, was recently raised by the suggested
gauge/string correspondence \cite{GKP,FroTse}.} The Kerr orientifold string
takes the form of an open D-string with joined ends which propagate along
the Kerr ring.

Our treatment is based on
the previous papers \cite{Bur-nst,BurAli} where the real and complex
structures of the Kerr geometry were considered. For the reader convenience
 we describe briefly the necessary details of these structures.

\section{The structure of the Kerr source and micrigeon with spin}

We use the Kerr-Schild approach to the Kerr geometry \cite{DKS},
which is based on the Kerr-Schild form of metric
\be g_{\m\n} = \h_{\m\n} + 2 h
k_{\m} k_{\n}, \label{ksa} \ee where
$ \h_\mn $ is metric of auxiliary
Minkowski space-time, $ h= \frac {mr-e^2/2} {r^2 + a^2 \cos^2 \theta},$ and
$k_\m$ is a twisting null field, which is tangent to the Kerr principal null
congruence (PNC) and is determined by the form \fn{The rays of the Kerr PNC
are twistors and the Kerr PNC is
determined by the Kerr theorem as a quadric in projective twistor space
\cite{Bur-nst}.}
 \be k_\m dx^\m = dt +\frac z r dz + \frac r {r^2 +a^2} (xdx+ydy) - \frac a
{r^2 +a^2} (xdy-ydx) .  \label{km} \ee The form of the Kerr PNC is shown on
Fig. 1.  It follows from (\ref{ksa}) that the field $k^\m$ is null with
respect to $\h_\mn$ as well as with respect to the full metric $g_\mn$, \be
k^\m k_\m = k^\m k^\n g_\mn =  k^\m k^\n \eta_\mn. \label{kgh} \ee

\begin{figure}[ht]
\centerline{\epsfig{figure=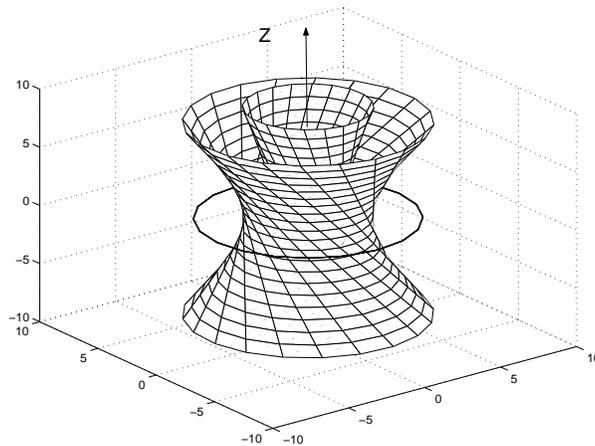,height=6cm,width=8cm}}
\caption{The Kerr singular ring and 3-D section of the Kerr principal null
congruence. Singular ring is a branch line of space, and PNC
propagates from ``negative'' sheet of the Kerr space to ``positive '' one,
covering the space-time twice. } \end{figure}

The metric is singular at the ring $r=\cos\theta=0$, which is
the focal region of the oblate spheroidal coordinate system $r, \theta, \phi$.

  The Kerr singular ring is the branch line of the
Kerr space on two folds:
positive sheet ($r>0$) and `negative' one ($r<0$).
Since for $|a|>>m$ the horizons disappear,
there appears the
problem of the source of Kerr solution with the alternative:
either to remove
this twofoldedness or to give it a physical interpretation.  The both
approaches have paid attention, and it seems that the both are valid for
different models.
The most popular approach was connected with the truncation
of the negative sheet of the Kerr space, which leads to the source
in the form
of a relativistically rotating disk \cite{Isr} and to the class of the
disk-like \cite{Lop} or bag-like \cite{Bag} models of the Kerr spinning
particle.

 Alternative way is to retain the negative sheet treating it as the sheet
of advanced fields. In this case the source of spinning particle
turns out to be the Kerr singular ring with the electromagnetic
excitations in the form of traveling waves, which generate spin and mass of
the particle.  Model of this sort was suggested in 1974 as a model of
`microgeon with spin' \cite{Bur0}. Singular ring was considered as a
 waveguide providing a circular propagation of an electromagnetic or
fermionic wave excitation. Twofoldedness of the Kerr geometry admits the
integer and half integer excitations  with $n=2\pi a/\lambda$ wave periods
on the Kerr ring of radfius $a$, which turns out to be consistent with the
corresponding values of the Kerr parameters $m= J/a$.

The light-like structure of the Kerr ring world-sheet is seen
from the analysis of the
Kerr null congruence near the ring. The light-like rays of the Kerr PNC are
tangent to the ring (see Fig.2).
\begin{figure}[ht]
\centerline{\epsfig{figure=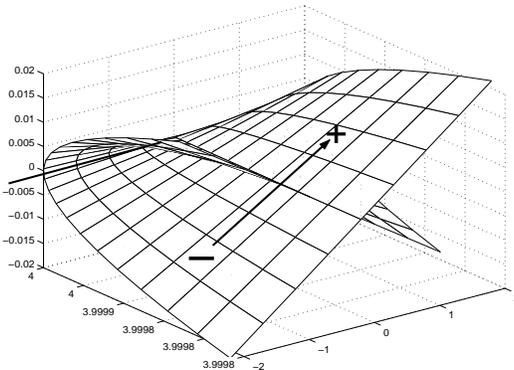,height=5cm,width=7cm}}
\caption{The (unoriented) surface $\phi=const.$ formed by the light-like
generators of the Kerr principal null congruence (PNC). The Kerr string (fat
line) is tangent to this surface.} \end{figure}

Formally it follows from the Eq.
(\ref{km}).  Approaching the ring ($r\to 0$) PNC takes the form
$ \breve k =k |_{r=\cos \theta=0}= dt - (xdy-ydx)/a = dt -a d\phi $,
and the light-like vector field $k_\m$ is tangent to the world
sheet of the Kerr ring. One sees that the Kerr ring is sliding along itself
with the speed of the light. \fn{It follows also from all the previous
treatments of the Kerr source, for example in \cite{Isr,Bur0,Lop}.
In spite of
the fact that function $h$ is singular at the ring, one can check that it is
true also with the respect to the auxiliary Minkowski metric, since in the
limit $r\to 0$, $ k^\n \breve k^\m g_{\mn}= k^\m \breve k^\n \eta _\mn \to
0.$}

\section{Stringy Nature of the Kerr Source}

It was recognized long ago \cite{IvBur} that the Kerr singular ring
 can be considered in the Kerr spinning particle as a string with
traveling waves.  We
recall here one of the most convincing evidences obtained by the analysis of
the axidilatonic generalization of the Kerr solution (given by \cite{Sen})
 near the Kerr singular ring .  It was shown \cite{BurSen} that the fields
 near the Kerr ring are very similar to the field around a heterotic string.

The limiting form of the Kerr-Sen metric near the ring is
\begin{equation}
ds_d^2 = e^{-2(\Phi -\Phi_0)}(dv^2 + du^2) + dy^2 - dt^2  +
(2Mr/\Sigma_d) (dy -dt)^2,
\end{equation}
and the gauge field in this limit is given by the vector potential
\begin{equation}
A =  2 Q (r/\Sigma_d)(dt - dy).
\end{equation}
By introducing an electric charge per unit length of the Kerr ring $ q=
2^{(3/2)}Q/(2 \pi a)$ and a two-dimensional (two-valued) Green's function
$G_a^{(2)}$ in the $(u,v)$ complex plane near the Kerr singularity
\begin{equation}
 G_a^{(2)}= 2 \pi ar/\Sigma \simeq  \pi \{[\frac{a}{2(u+iv)}]^{1/2} +
  [\frac {a}{2(u-iv)}]^{1/2}\},
\end{equation}
the dilaton factor may be represented as
\begin{equation}
e^{-2(\Phi -\Phi_0)} = 1 + N G_a^{(2)} ,
\end{equation}
where
$N =r_- /2\pi a.$

Then the rescaled $\sigma$-model metric
$ ds_{str}^2 =  e^{2(\Phi -\Phi_0)}ds_d^2, $
used in string theory may be written in the form
\begin{equation}
ds_{str}^2 = (dv^2 + du^2) +
\frac{1}{1+ NG_a^{(2)}}
(dy^2 - dt^2) + \frac{2M G_a^{(2)}}{2\pi a(1+ NG_a^{(2)})^2} (dy -dt)^2
\end{equation}
which is exactly  the form of metric obtained by Sen
for a field around a fundamental heterotic string \cite{Sen0}.
\par
The difference is only in the form of two-dimensional
Green's function $G_a^{(2)}$, but it is very natural and caused
by the twovaluedness of the fields near the Kerr singularity.

\section{The Kerr Orientifold World-Sheet}

One can see that the world-sheet of
the Kerr ring satisfies the bosonic string equations and
constraints, however, there appear the problems with boundary conditions.

 The general solution of the string wave equation
$(\frac {\d ^2} {\d \sigma ^2} - \frac {\d ^2} {\d \t ^2}) X^\m  =0$
can
be represented as the sum of the `left' and `right' modes:
$X^\m(\sigma, \t)=
X_R^\m (\t -\sigma) + X_L^\m (\t +\sigma), $ and the oscillator expansion is
\be X_R^\m (\t -\sigma) =\frac 12 [ x^\m + l^2 p^\m (\t -\sigma) +
il\sum_{n\ne0} \frac 1n \alpha_n^\m e^{-2in(\t-\sigma)}] ,
\label{R} \ee

\be X_L^\m (\t +\sigma) =\frac 12 [ x^\m + l^2 p^\m (\t +\sigma) +
il\sum _{n\ne0} \frac 1n \tilde\alpha_n^\m e^{-2in(\t+\sigma)}] ,
\label{L} \ee

where $ l =\sqrt{2\alpha^\prime} =\frac 1{\sqrt{\pi T}}$,  $T$ is tension,
$x^\m$ is position of center of mass, and $p^\m$ is momentum of string.

The string constraints
$\dot X_\m \dot X^\m + X^{\prime}_\m X^{\prime\m} =0, \qquad
\dot X_\m  X^{\prime\m} =0,$
are satisfied if the modes are light-like ($()^\prime \equiv \d _\sigma ()$),
\be
(\d _ \sigma X_{L(R)\m}) (\d _{\sigma} X_{L(R)}^\m) =0. \label{c1}
\ee
Setting $2\sigma = a \phi$ one can describe the light-like world-sheet of
the Kerr ring (in the rest frame of the Kerr particle) by the surface
\be
X_L^\m(t,\sigma) = x^\m + \frac 1{\pi T} \delta _0^\m p^0 (t + \sigma) +
\frac
a 2 [(m^\m +in^\m) e^{-i2(\t+\sigma)} + (m^\m -in^\m) e^{i2(\t+\sigma)}]  ,
\label{kring} \ee
where
$m^\m$ and $n^\m$ are two space-like basis vectors lying in the plane of the
Kerr ring.  One can see, that
\be X_L^{\prime\m} =
\frac 1{\pi T} \delta _0^\m  p^0  + 2a[-m^\m \sin 2(\t+\sigma) + n^\m \cos
2(\t+\sigma)] \label{dkring} \ee will be a light-like vector if one sets $p^0
= 2\pi a T $.
It shows that the Kerr world-sheet could be described by modes of one (say
`left') null direction. The solution $X(\t, \sigma) =X_L(\t+ \sigma)$
 satisfies the string wave equation and constraints, but
there appears the problem with boundary conditions.
The closed string boundary condition
\be X^\m(\t,\sigma)=X^\m(\t,\sigma+\pi)
\label{cl-b} \ee
 will not be satisfied since the time component
$X_L^0 (t,\sigma +\pi)$ acquires contribution from the second term in
(\ref{kring}), which is usually compensated by this term from the
`right' mode.
The familiar boundary conditions for the open strings follow from the
condition of canceling of the surface term
$ -T\int d\t [X^\prime _\m \delta X^\m |_{\sigma=\pi} -
X^\prime _\m \delta X^\m |_{\sigma=0} ]$ in the string action  \cite{GSW},
and are
\be
X^{\prime \m}(\t,0)= X^{\prime \m}(\t, \pi)=0,
\label{op-b} \ee
which also demand the both types of modes to form standing waves.
However, this demand can be weakened  to
\be
X^{\prime \m}(\t,0)= X^{\prime \m}(\t ,\pi).
\label{Kop-b} \ee
It seems that the light-like oriented string can contain traveling
waves of only one direction if we assume that it is open, but has
the joined ends.
However, the ends  $\sigma=0,$ and $\sigma=\pi$ are not
joined  indeed.

These difficulties can be removed by the formation of the world-sheet
orientifold.

It is well known \cite{GSW} that the interval of an open string  $\sigma
\in [0,\pi] $ can be formally extended to $[0,2\pi]$, setting
\be X_R (\sigma
+\pi) =X_L (\sigma), \qquad X_L (\sigma +\pi) =X_R (\sigma).  \label{ext} \ee
By such an extension, the both types of modes, `right' and `left', will
appear in our case since the `left' modes will
play the role of `right' ones on the extended piece of interval.
If the extension is completed by the changing of orientation on the
extended piece,  $\sigma ^\prime = \pi - \sigma $, with a subsequent
identification of $\sigma$ and  $\sigma ^\prime$, then one obtains
the closed string on the interval $[0,2\pi]$ which is
 folded and takes the form of the initial open string.

Formally, the world-sheet orientifold
represents a doubling of the world-sheet with the orientation reversal on the
second sheet. The fundamental domain $[0,\pi]$ is extended to
$\Sigma=[0,2\pi]$ with formation of folds at the ends of the interval
$[0,\pi]$.

\begin{figure}[ht]
\centerline{\epsfig{figure=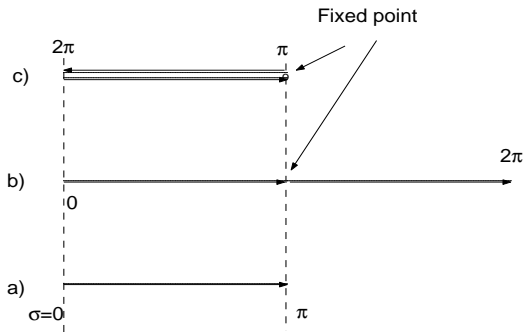,height=5cm,width=7cm}}
\caption{Formation of orientifold: a) the initial string interval, b)
extension of the interval and formation of the both side movers, c)
formation of the orientifold.} \end{figure}

The parity operator $\Omega$ : $ \sigma \to 2\pi - \sigma$
changes the layers of the world-sheet.
The ends of the fundamental domain $\sigma =0$ and $\sigma=\pi$ are the fixed
points of $\Omega$.  Applying the operator
$\Omega$ to (\ref{ext}), one sees that $\Omega$ acts on the string by changing
the right and left modes, and as a result the solution
$X= X_R + X_L$ turns out to be parity invariant.
Orientifold is the factor space $\Sigma/\Omega$ setting the equivalence of
the original interval and the extended piece.

\begin{figure}[ht]
\centerline{\epsfig{figure=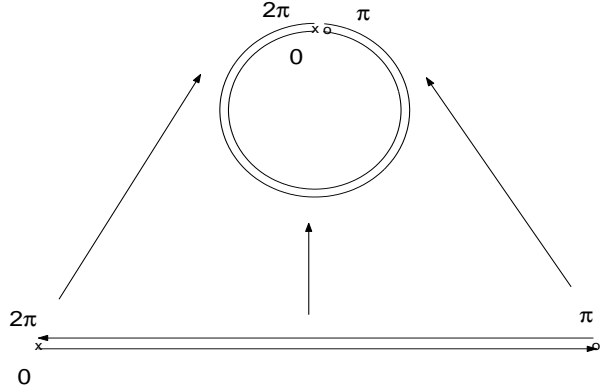,height=6cm,width=8cm}}
\caption{Formation of the Kerr D-string.}
\end{figure}

Orientifolding
 the Kerr light-like world-sheet  $X=X_L(\t+\sigma)$, containing only the
left modes on the fundamental interval $[0,\pi]$, one obtains the open
string which can be represented as closed one on the covering space, but
folded taking the form of open string. Solution
$X(\t,\sigma)=X_L(\t+\sigma)+X_R(\t-\sigma)$ contains the both types of modes,
and the discussed above linear in $\sigma$ term cancels. As a result
the ends of the open string turns out to be joined.

The resulting  orientifold string retains
the form of the Kerr light-like string which is covered twice in opposite
directions.  The joined ends of the string are in the light-like circular
motion along the string.

\section{Some Relations to D-branes and Superstring Theory}

By orientifolding the Kerr
string, it acquires simultaneously the properties of the open and closed
strings. Being formally closed, it forms a configuration with open ends
joined to the fixed points of the operator $\Omega$.
With the respect to the initial model of
the Kerr geon, the Kerr spinning particle acquires some new features connected
with the light-like orientifold fixed points, which are the
peculiar points of the model.  The orientifold is
closely related to D-branes which are carriers of the RR-charges
\cite{Pol}.
The ends of the open string has to be stuck to D-branes
and supplied by Chan-Paton factors corresponding to some gauge group of
 symmetry G  \cite{Pol,GimPol,GSW}.\fn{Nice introduction is given in
\cite{Joh}.}    In superstring
theories p-brane is the brane with p+1 Neymann boundary conditions \cite{Pol}
and Dirichlet conditions on the remaining 9-p of the 10 dimensional
space-time.  Looking on the boundary conditions (\ref{Kop-b}) one sees that
they are Neumann conditions along the directions tangent to the Kerr string
(including time), and Dirichlet conditions along directions transverse to the
string.  Therefore, the Kerr string can be considered as a Dp-brane with p=1,
or the D-string of the type I string theory.  It was shown in
\cite{Dab,Hul,PolWit} that worldsheet of the D-string has the structure of a
heterotic string, which is responsible for the considered in sec.III heterotic
properties of the Kerr string.

 On the other hand, the type I string theory can
be obtained orientifolding the type IIB string theory \cite{PolWit,Joh},
which requires the presence of Dirichlet 9-branes as the end points of the
string \cite{Hull9705,Hull9812}.\fn{A 9-brane fills 9 dimensional space of 10
dimensional space-time, so they do not restrict the motion of the end points
of the string, but form the Chan-Paton degrees of freedom \cite{Pol}.  The
type I theory by compactification on K3 contains D5-branes. A pair of
D5-branes carry SU(2) Chan-Paton factors
\cite{WitIns,DabPar,GimJoh,GimPol,Sen9805}.}
Since D-branes are oriented, and the
D-branes of opposite orientation, anti-D-branes \cite{BanSus}, carry
opposite charge, there appears the possibility to form the
  D-string anti-D-string pairs which are not BPS states
\cite{Sen9904},  but are the lightest stable states
\cite{Sen9803,Sen9805,Sen9904}. It represents especial interest since the
Kerr solution with the parameters of the elementary particles is very far
from the BPS one.

The obtained orientifold structure
of the Kerr string displays some new features of the Kerr spinning particle.
In particular, the light-like open Kerr string with the Chan-Paton
factors  resembles the well known model of the light-like quarks
sitting on the ends of a gluonic string.

\section{The Kerr spinning particle and the Complex Kerr string }

We have to discuss excitations of the Kerr D-string. The mixed type of the
boundary conditions (Dirichlet-D and Neumann-N) yield three sectors of the
excitation spectrum of the D-string. It was studied in \cite{Hul,Dab,PolWit}
and shown to reproduce the world sheet structure of the heterotic string.
The DD-sector of excitations yields the right-handed world-sheet fermions
and  scalars leading to oscillations in position of the D-string
\cite{PolWit}. With application to the Kerr string it corresponds to
oscillations in the form and position of the string. The general dynamics of
the Kerr singular ring is determined by the Kerr theorem \cite{Bur-nst}.  The
recent progress in the obtaining of the nonstationary Kerr solutions  is
connected with a complex representation of the Kerr geometry \cite{Bur-nst}.
The nonstationary solutions can be represented in this approach as the
retarded-time fields.  However, in the Kerr case they are generated by a {\it
complex} source moving along a {\it complex world line} $x_0(\t)$ in complex
Minkowski space-time $CM^4$.  The real part of the complex world line controls
the position of the Kerr ring, and the imaginary part determines orientation
of the Kerr singular ring, connected with spin \cite{Bur-nst,BurAli}.

 The objects described by the complex world lines occupy an intermediate
position between particle and string.  Like the string they form the
two-dimensional
surfaces or the world-sheets in the space-time \cite{OogVaf,BurStr}. The
corresponding complex Kerr string in many respects is similar to the
"mysterious" $N=2$ complex string of superstring theory \cite{OogVaf}, but it
has the euclidean world sheet. It was shown in  \cite{BurStr}
that the boundary conditions of the complex Kerr string demand also the
orientifold structure,  similar to the Kerr singular ring.
Moreover, the corresponding complex string equations admit the analytic in
$\t$ solutions which are analogues to the above one side stringy movers, and
this complex orientifold restores the movers of both
sides.\fn{Note, that orientifold was introduced by Sagnotti and Horava
in 1989 \cite{SagHor}, and was called by Horava the world-sheet orbifold.
In the papers \cite{BurStr} it was independently considered by analysis of the
complex Kerr string. After electronic publication \cite{BurStr},
Horava informed us on his works. The term orientifold has gained the
broad currency after the well-known works by
Polchinski and Witten \cite{Pol,PolWit}.}

Let us now discuss the origin of the NN and DN excitations.
The NN-sector in which both ends satisfy the Neumann boundary conditions
carries the Chan-Paton factors leading to the usual open strings of type I
string theory. This sector can be interpreted in the Kerr case as a
fundamental (1,0) string coupled to the (0,1) D-string, forming a (1,1)
system. It corresponds in the microgeon model to the role of the Kerr D-string
 as a waveguide keeping the traveling e.m. waves.

Excitations of DN sector are responsible
 for the current algebra modes. The structure of this sector is represented
 as a cloud of the fundamental DN-strings surrounding the D-string
 \cite{Hull9812}. The D-ends of these strings are stuck to the D-string and
 the N-ends are free (coupled to D9-branes) and carry the Chan-Paton factors.
This sector acquires a semiclassical explanation in the model of Kerr
spinning particle.
The oscillating Kerr solutions are classically radiative
\cite{Bur-nst,BurAli}, and the null rays of the  Kerr PNC are the lines of
propagation of the electromagnetic radiation (pp-waves in the Penrose limit).
The stress energy tensor of the radiation has the form $T_{rad}^\mn=
\Phi k^\m k^\n$, where $\Phi \sim 1/r^2$, and $k^\m$ is
tangent to the Kerr PNC. Therefore, radiation leads to an infrared divergence
of the total mass.  However, the Kerr radiation has the remarkable property
that its flow is conserved $\nabla _\m T_{rad}^\mn =0$ ,
and  can be decoupled from the Kerr singular source. Therefore, the divergence
can be `regularized' by the simple subtraction of the radiative terms
\cite{Bur-nst,BurAli}. It can also be seen from the Fig.1 since the flow
of the out-going radiation has extension to the Kerr negative sheet of
advanced fields where it turns into in-going field. A very small part of the
rays of PNC lying in the equatorial Kerr plane $\theta =\pi/2$ reaches the
Kerr D-string, but it does not lead to the above
divergence.  These rays are carriers of the fundamental pp-waves
which are responsible for the DN-sector of the Kerr D-string. \fn{From this
point of view the DN excitations of the Kerr D-string represent a resonance 
of the e.m. zero-point field on the Kerr D-string, which is an one 
dimensional version of the Casimir effect.}

The dynamics of the Kerr ring is closely related with its e.m.
excitations and determined by the interaction of the real and complex Kerr
strings.  We refer reader for details to the papers
\cite{BurStr,Bur-nst,BurAli}.  The development  of this approach is connected
with obtaining  the selfconsistent oscillating solutions of the corresponding
Einstein-Maxwell system of equations, and with a further generalization of
these solutions to supergravity.  It is a very hard problem.  The obtained
recently class of nonstationary Kerr solutions \cite{Bur-nst} represented some
progress in this direction.  The preliminary attempts to get selfconsistent
solutions \cite{BurAli} have showed  that oscillations  lead to the appearance
of the imaginary mass term (dual mass), which is similar to the NUT parameter
\cite{KraSte}, but is also oscillating in this case.

\section*{Acknowledgments}

We are thankful to the participants of the Workshop ``Supersymmetry and
Quantum Symmetry'' (JINR, Dubna, July 2003) for very useful discussion, in
particular to A. Zheltukhin, I. Bandos, A. Tseytlin and G. Alekseev.

\end{document}